# Human-machine cooperation: optimization of drug retrieval sequencing in automated drug dispensing systems

Mengge Yuan[a] Kan Wu[b] and Ning Zhao[a]*

[a]Faculty of Science, Kunming University of Science and Technology, Kunming, Yunnan; [b]Business Analytics Research Centre, Chang Gung University, Taoyuan City, Taiwan

Corresponding author: Ning Zhao; E-mail: zhaoning@kust.edu.cn; Faculty of Science, Kunming University of Science and Technology, Kunming, Yunnan

# Human-machine cooperation: optimization of drug retrieval sequencing in automated drug dispensing systems

**Abstract**: Automated drug dispensing systems (ADDSs) are increasingly in demand in today's pharmacies, primarily driven by the growing ageing population. Recognizing the practical challenges faced by pharmacies implementing ADDSs, this study aims to optimize the layout design and sequencing issues within a human-machine cooperation environment to enhance the system throughput of ADDSs. Specifically, we develop models for drug retrieval sequencing under different system layout designs, taking into account the stochastic sorting time of pharmacists. The prescription order arrival pattern follows a successive arrival mode. To assess the efficiency of ADDSs with one input/output point and two input/output points, we propose dual command retrieval sequencing models that optimize the retrieval sequence of drugs in adjacent prescription orders. Notably, our models incorporate the stochastic sorting time of pharmacists to analyze its impact on ADDS performance. Through experimental comparisons of average picking times for prescription orders under various operational conditions, we demonstrate that a system layout design incorporating two input/output points significantly enhances the efficiency of prescription order fulfilment within a human-machine cooperation environment. Furthermore, our proposed retrieval sequencing method outperforms dynamic programming, greedy, and random strategies in terms of improving prescription order-picking efficiency. By addressing the layout design and sequencing challenges, our research contributes to the field of intelligent warehousing, particularly in smart pharmacies. The findings provide valuable insights for healthcare facilities and organizations seeking to optimize ADDS performance and enhance drug dispensing efficiency.

## 1. Introduction

The increasing demand for automated drug dispensing systems (ADDSs) in smart pharmacies necessitates enhanced throughput to meet the needs of the growing ageing population. The ADDS, a highly efficient human-machine coordinated prescription order picking system, can increase pharmacy picking efficiency and shorten pharmacists' walking distance (Azadeh et al., 2019 ; da Costa Barros & Nascimento, 2021 ; Zhang et al., 2023), as shown in Fig. 1. The order-picking problem has long been identified as the most labor-intensive and costly activity for almost every warehouse and the ADDS has gained significant importance in smart pharmacies due to enhanced prescription order-picking processes. The ADDS carries out automatically various tasks such as receiving, storing, picking, and dispensing, leading to increased efficiency, accuracy, and patient safety, compared to traditional pharmacies (Lin & Hsieh, 2017). Among these activities, order picking stands out as the costliest process, involving the retrieval of drugs from their designated storage locations (De Lombaert et al., 2022). The ADDSs face challenges in effectively handling the increasing number of prescription orders, resulting in delays and longer waiting times for patients (Khader et al., 2016 ; van Gils et al., 2018). This situation calls for innovative solutions to optimize the throughput of smart pharmacies. Improving the throughput in smart pharmacies is crucial to ensure timely and efficient drug dispensing, reducing patient wait times, enhancing workflow efficiency, and optimizing resource utilization.

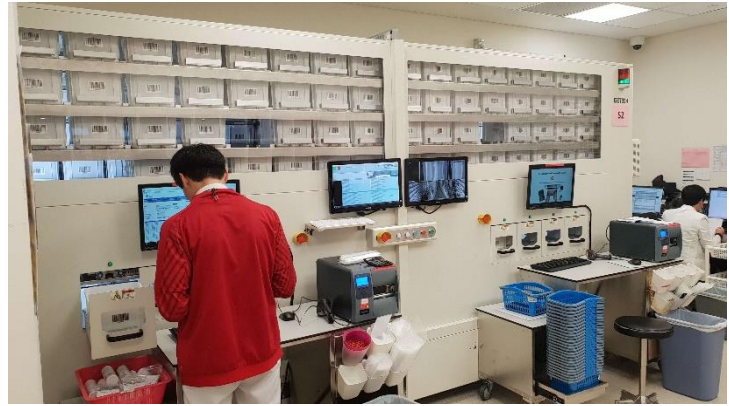

**Fig. 1.** Visual representation of an ADDS at Changi General Hospital, Singapore.

ADDSs collaborate closely with order pickers to fulfil orders efficiently, as shown in Fig. 1. It's worth noting that ADDSs share similarities with unit-load automated storage and retrieval systems (AS/RSs) (Salah et al., 2017). The main components of an AS/RS are racks, cranes, aisles, input/output (I/O) points, and pick positions. These handling systems can be aisle captive (typically cranes). An important decision problem for efficiently operating an AS/RS is the crane scheduling problem, which decides on the sequence of crane moves for processing specific storage and retrieval requests (Boysen & Stephan, 2016).

However, different from AS/RS, ADDSs often need to meet specific requirements in the pharmaceutical industry, such as ensuring drug safety and precision. ADDSs demand a high level of accuracy in drug distribution since the dosage and type of drug significantly impact patient treatment outcomes and safety. Therefore, these systems rely on the expertise of pharmacists to assist in precise drug sorting and packaging.

Consequently, it is crucial to conduct in-depth research on optimizing the design of systems like ADDSs.

For optimizing the design of ADDSs, we focus on the physical design and control aspects of an order-picking problem because they form the heart and soul of any warehouse (Azadeh et al., 2019). Motivated by De Koster et al. (2007) and Roodbergen and Vis (2009), a simple schematic view of design issues and their interdependence is presented, as depicted in Fig. 2. According to the author's understanding, on the one hand, the performance of ADDSs in smart pharmacies is intricately connected to the number of I/O points for layout design within the system. Given the substantial cost involved in implementing ADDSs, it becomes imperative to conduct comprehensive research at the outset, exploring the influence of different I/O point quantities on system performance, and providing invaluable recommendations to healthcare facilities. On the other hand, of particular interest to us is to sequence the drug bins on the pick list to ensure a good route for prescription order picking. How can we enhance drug retrieval efficiency? These are the thought-provoking questions that captivate our attention. Therefore, we focus on the number of I/O points and the sequencing method in ADDSs, which play an important role in improving the throughput of the system (Azadeh et al., 2019 ; Chen et al., 2023 ; Gu et al., 2010).

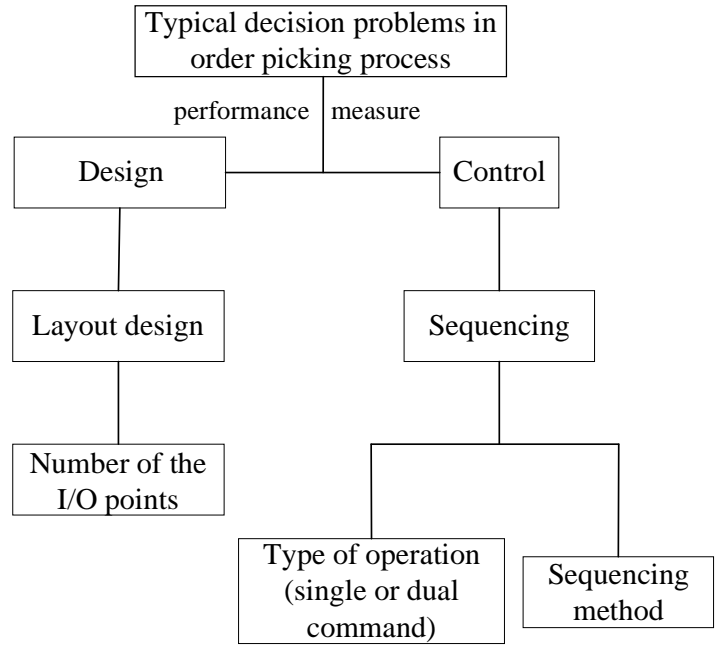

**Fig. 2**. The design of an ADDS.

The crane of ADDSs can work under two main operating types: single and dual command cycles, which also have an impact on the sequencing. In a single command (SC) cycle, the crane performs either a single storage or a single retrieval request in a cycle. If the crane performs both a storage and a retrieval request in a cycle, it is referred to as a dual command (DC) cycle, as shown in Fig. 3. Comparatively, the dual command cycle outperforms the single command cycle in reducing the time required for storage and retrieval requests (Yu & De Koster, 2012 ; Zhen & Li, 2022). Pohl et al. (2009) and Bortolini et al. (2020) optimized the number and location of aisles in a warehouse using dual-command operations. Therefore, the operational way the ADDSs perform dual command operations whenever possible, and single command operations otherwise.

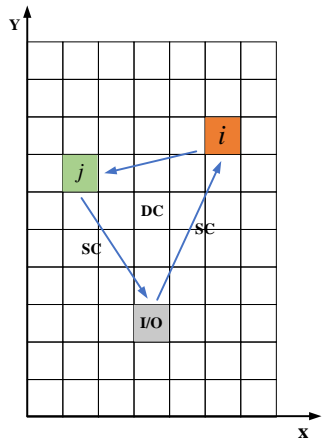

**Fig. 3.** Single command cycle and dual command cycle operations.

In smart pharmacies, the application of a progressive zone (PZ) strategy for picking prescription orders has proven advantageous for smaller-scale order fulfilment (Azadeh et al., 2023). Under the PZ strategy, the pharmacy is divided into multiple zones, with each zone housing an ADDS. One or multiple pharmacists are assigned to each machine and exclusively perform picking tasks within their designated zone, ensuring a collaborative environment between the ADDS and pharmacists for efficient and ergonomically friendly order fulfilment (Azadeh et al., 2019). As depicted in Fig. 4 (da Costa Barros & Nascimento, 2021), within this collaborative setting, the ADDS autonomously handles the storage and retrieval of drug bins, while pharmacists are responsible for manually sorting the required dosage at the I/O point. Together, they fulfil all the necessary pick-list items within that specific zone. If an order remains incomplete, the pharmacist manually pushes the pick cart, containing the partially fulfilled items, to the pharmacist in another zone. Conversely, if all required drugs are successfully picked, the pharmacist in the last zone manually pushes the pick cart to the dispensing counter, making the machine available for processing the next order. In PZ picking, additional order consolidation is unnecessary since the orders are gradually consolidated as the order visits each zone. By allowing ADDS to primarily handle drug retrieval, the non-productive walking time of pharmacists is reduced, thereby enhancing picking efficiency. Our research focus lies on improving the overall picking performance between humans and machines within each zone.

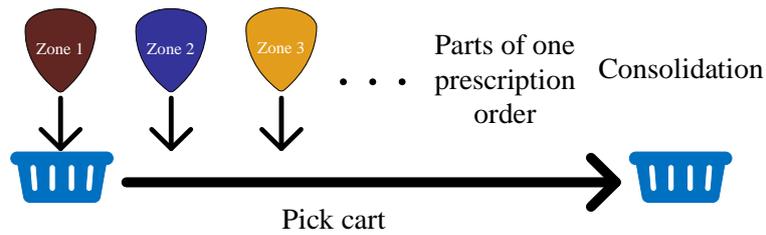

**Fig. 4.** Progressive zone strategy.

The prime contributions of this paper are as follows: (1) we are the first to consider the pharmacists' stochastic sorting time in the order-picking problem for ADDSs, (2) we develop models to analyze the picking performance of ADDSs with two different system layout designs: one I/O point layout and two I/O points layout, (3) we optimize the drug retrieval sequencing method in the scenario of successive order arrivals to improve the overall order picking performance.

The rest of the study is organized as follows. Section 2 reviews the related literature. Section 3 provides the system description of ADDSs. In Section 4, retrieval sequencing models are developed for various system layout designs. Section 5 conducts numerical experiments in two parts: the performance comparison of ADDSs with different system layout designs and drug retrieval sequencing strategies comparisons. Finally, Section 6 gives the conclusion and provides an outlook for future research.

**2. Literature Review**

The operational performance of ADDSs can generally be improved with layout design and control rules. Despite the fact that decisions in both categories are highly interrelated, they are usually addressed separately (Roodbergen & Vis, 2009). In this section, we focus on the literature about the number of I/O points for the design issue, the sequencing method for the control issue and the order-picking problem under the human-machine cooperation environment.

**Layout design:** For the physical design in ADDSs, we focus on the number of input/output (I/O) points, which hardly any attention has been paid to (Randhawa et al., 1991 ; Randhawa & Shroff, 1995). The majority of the literature on AS/RSs provides design and operational choices assuming a single I/O point for the system. However, many systems have multiple I/O points like ADDSs. New studies are required to investigate questions such as, what is the effect of having multiple I/O points on the design and operational choices (Azadeh et al., 2019 ; Roodbergen & Vis, 2009). This paper aims to study the system performance of ADDSs with one I/O point design versus two I/O points design, which may have an impact on the throughput.

**Sequencing:** Sequencing rules can be used to create tours such that the total time to handle all requests is minimized or the due times are least violated (Roodbergen & Vis, 2009). The sequencing problem, as demonstrated in Mahajan et al. (1998), is known to be NP-hard. Mahajan et al. (1998) improved the system throughput by appropriately sequencing retrieval requests within an order and optimizing retrieval requests among successive orders. Yu and De Koster (2012) studied how to sequence a group of storage and retrieval requests in a multi-deep automated storage system to minimize the makespan. Chen et al. (2015) discussed the integrated order batching, sequencing and routing problem in warehouses. Wauters et al. (2016) proposed a general mathematical model and efficient branch and bound procedure for optimizing the sequence of an AS/RS equipped with a dual shuttle crane. Ardjmand et al. (2018) developed a mathematical model to minimize order picking makespan in a warehouse of a major US third-party logistics company and the warehouse studied employs traditional wave picking with multiple pickers. For sequencing the release of bins from the ASRS, Boysen et al. (2018) derived an elementary optimization problem to minimize the spread of orders in the release sequence so that picking orders are quickly assembled at their packing stations. Yang et al. (2021) addressed the problem of joint optimization of order sequencing and rack scheduling in the robotic mobile fulfilment system to improve system performance. Yang et al. (2023) studied the optimal sequence of the joint shuttle transfer and load retrieval, such that the makespan of a set of requests was minimized. For an innovative automation system like ADDS, we are also interested in how to sequence drug requests for prescription order-picking problems.

**Human-machine cooperation:** By deploying humans to complement machine labor or by using machines to supplement human labor, humans still play a key role in pharmacy operations. De Lombaert et al. (2022) and Pasparakis et al. (2023) demonstrated that traditional order picking is being re-designed into collaborative human-robot tasks and this trend exemplifies the transition towards a human-centric Industry 5.0, focusing on synergy instead of seeking a human replacement. Wang et al. (2022) studied the problem of finding a suitable robot schedule that considers the schedule-induced fluctuation of the working states of human pickers and proposed a model to minimize the expected total picking time in human-robot coordinated order-picking systems. Löffler et al. (2023) treated the coordination of multiple AMRs and multiple pickers to minimize the makespan and analyzed the effects of stochastic picking times and speed differences between AMRs and pickers. Azadeh et al. (2023) studied the effect of zoning strategies on throughput capacity and reduced the pickers' unproductive walking time for human-robot collaborative picking. In the same way, ADDSs operate within a human-machine cooperation environment. ADDSs can handle the drug retrieval requests, while some other works, such as the sorting of pills, tablets, and Chinese herbs at the I/O point still require intervention by pharmacists (Li et al., 2022). This human sorting process at the I/O point is time-consuming and can significantly impact the system's operations and efficiency. The interesting research question is, how can we coordinate the parallel working of the pharmacists and ADDSs to maximize the picking efficiency? However, limited attention has been given to the specific issue of drug retrieval sequencing among prescription orders and its impact on system efficiency in a human-machine cooperation environment.

Integrating layout design and sequencing problems can yield substantial efficiency benefits, which are crucial in adapting to market developments (van Gils et al., 2018). Motivated by practical needs, our research explores various strategies for optimizing the operations of ADDSs, including the operation modes of cranes, drug retrieval sequencing, and layout designs of systems. We focus on investigating drug retrieval sequencing among successive prescription orders and analyzing the impact of stochastic service time of pharmacists on ADDS performance. Also, we formulate retrieval sequencing models for various system layout designs. Finally, we conduct a series of numerical experiments to evaluate the efficiency of the ADDS, taking into account the diverse layout designs and varying parameters related to human factors. The proposed retrieval sequencing method is a superior approach compared to existing strategies such as dynamic programming, greedy, and random methods.

## 3. System Description

In this section, we introduce the details of ADDS including the basic structure, the workflow, and the order arrivals.

### 3.1. Basic structure

The ADDS consists of a double-sided storage rack, a crane, a robotic arm, a track, and I/O points (Yuan et al., 2023). The storage locations on both sides of the machine are considered grid planes (refer to Fig. 5). Each storage location accommodates a drug bin. Each bin stores only one type of drug and one type of drug may be stored in multiple bins. The robotic arm runs on the middle track and can pick up any bin on both sides. It moves simultaneously in both horizontal and vertical directions at a constant speed $v$, with negligible acceleration and deceleration.

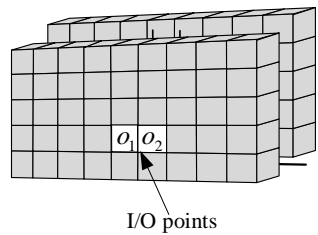

**Fig. 5.** The system diagram for the ADDS.

Each bin is located on the storage rack and its location is denoted by $i$, with coordinates $(x_i, y_i)$, where $x_i$ is the row, $y_i$ is the column of the storage location. Additionally, the height and length of each bin are represented by $d_m$ and $d_n$, respectively. Let $(x_0, y_0)$ denote the coordinates of the I/O point. The travel time from location $i$ to the I/O point is represented by $\max\{|x_i - x_0| \times d_m/v, |y_i - y_0| \times d_n/v\}$. The dual command travel time, $t_{i,j}$, can be expressed by the following equation (1) for the route $(I/O)O \to i \to j \to O(I/O)$ in Fig. 3. Assuming the service time $X$ of pharmacists at the I/O point is normally distributed, i.e. $X : N(\mu, \sigma^2)$, where $\mu$ is the mean and $\sigma$ is the standard deviation.

$$t_{i,j} = \max\{|x_i - x_0| \times d_m/v, |y_i - y_0| \times d_n/v\} + \max\{|x_i - x_j| \times d_m/v, |y_i - y_j| \times d_n/v\} \\ + \max\{|x_j - x_0| \times d_m/v, |y_j - y_0| \times d_n/v\}. \tag{1}$$

### 3.2. The workflow of ADDSs

We introduce the workflow of ADDSs under two layout designs as follows: layout A and layout B.

**Layout A.** There are two I/O points on the storage rack of the ADDS, as shown in Fig. 5. Upon receiving a prescription order request, the robotic arm transports the retrieved drug bin to the I/O point $O_1$, and the time is $t_{i,j}$. Then the pharmacist sorts the required dosage of the drug bin at $O_1$ and the sorting time is $X$. Meanwhile, the robotic arm retrieves the next drug bin and brings it to the I/O point $O_2$. Then the pharmacist sorts the necessary dosage from the drug bin at $O_2$. The cooperation between the pharmacist and the ADDS is synchronized, and the picking time is $\max(X, t_{i,j})$. This means that while the ADDS retrieves one drug bin, the pharmacist can sort another type of drug at the I/O point. During the time interval $\max(X, t_{i,j})$, the pharmacist completes drug sorting at the I/O point and the robotic arm returns the drug bin to location $i$ and then retrieves a drug bin in location $j$ following the route $O_p \to i \to j \to O_p$, where $p \in \{1, 2\}$.

**Layout B.** For layout B, there is only one I/O point on the storage rack of the ADDS. After receiving the prescription order request, the robotic arm retrieves the drug bin from the storage rack and then sends it to the I/O point. Then the pharmacist retrieves the necessary dosage of the drug at the I/O point. After that, the robotic arm returns the bin which is handled by the pharmacist at the I/O point to its storage location and then retrieves the next drug request. This process is repeated until all drugs in the prescription are retrieved. The cooperation between the pharmacist and the robotic arm in layout B is a sequential operation. That is, the ADDS doesn't perform any operation until the pharmacist completes the service at the I/O point. So the total time that one drug is sorted at the I/O point and the robotic arm returns it to location $i$ and retrieves a drug bin in location $j$ following the route $O_1 \to i \to j \to O_1$ is $X + t_{i,j}$.

### 3.3. Order arrivals

Assuming ADDSs operate based on a first-come-first-served strategy for fulfilling prescription orders. There are two arrival scenarios: the successive scenario and the non-successive scenario. For the successive scenario, while a previous prescription order is still being processed and the I/O points are occupied, the information system of an ADDS has a new prescription order in the queue. This leads to an uninterrupted sequence of dual command operations. The retrieval sequence in this scenario is influenced by the adjacent prescription orders.

We provide an example to illustrate the successive scenario. Let $i_k$ denote the location of drug $k$. Suppose there are two prescription orders, where the retrieval sequence for order 1 is $i_1, i_2, \ldots, i_{k_1}$, and the retrieval sequence for order 2 is $i_1', i_2', \ldots, i_{k_2}'$, as shown in Fig. 6. Drugs $i_{k_1-1}$ and $i_{k_1}$ need to be returned by the robotic arm before retrieving a new order for ADDSs under layout A. Therefore, drug locations $i_{k_1-1}$ and $i_{k_1}$ influence the path of the new order picking. Under layout B, the drug location $i_{k_1}$ follows the same operation as described above.

Layout A: $i_1 \rightarrow i_2 \rightarrow \cdots \rightarrow \boxed{i_{k_1-1} \rightarrow i_{k_1}} \rightarrow i_1' \rightarrow i_2' \rightarrow \cdots \rightarrow i_{k_2}'$

Layout B: $i_1 \rightarrow i_2 \rightarrow \cdots \rightarrow \boxed{i_{k_1}} \rightarrow i_1' \rightarrow i_2' \rightarrow \cdots \rightarrow i_{k_2}'$

**Fig. 6.** Retrieval sequence for adjacent prescription orders under layout A and layout B.

In the non-successive scenario, there is a longer interarrival time between prescription orders. After fulfilling one prescription order, the robotic arm has enough time to return the drug bin from the I/O point to its storage location and then return to the I/O point, waiting for a new order. Thus, the picking process of prescription orders is independent and does not affect each other. We provide a picking path of an order under Layout A and Layout B respectively. Let $K$ represent the number of types of drugs in a prescription order, where $K \geq 2$. The picking route of layout A is $O_1 \rightarrow i_1 \rightarrow O_1$, $O_2 \rightarrow i_2 \rightarrow O_2$, $O_1 \rightarrow i_1 \rightarrow i_3 \rightarrow O_1$, $O_2 \rightarrow i_2 \rightarrow i_4 \rightarrow O_2$, L, $O_2 \rightarrow i_{K-2} \rightarrow i_K \rightarrow O_2, O_1 \rightarrow i_{K-1} \rightarrow O_1, O_2 \rightarrow i_K \rightarrow O_2$, as shown in Fig. 7. Four single command operations and $K-2$ dual command operations are needed. The picking route of layout B is $O \to i_1 \to O$, $O \to i_1 \to i_2 \to O$, $\mathsf{L}$, $O \to i_{K-1} \to i_K \to O$, $O \to i_K \to O$, where two single command operations and $K-1$ dual command operations are needed.

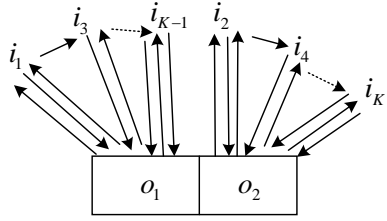

**Fig. 7**. Picking route in a non-successive scenario for layout A.

Note that non-successive scenarios only occur at the beginning and end of a busy period of an ADDS. The throughput of ADDSs is determined by their performance in successive scenarios. In this paper, we primarily focus on the modelling and analysis of the prescription order-picking problem under the assumption of successive order arrivals.

**4. Retrieval sequencing model**

In a human-machine cooperation environment, the picking sequence of $K$ drugs in a prescription order impacts the operational efficiency of the system. In this section, models are established for the successive arrival scenario of prescription orders under layout A and layout B respectively to optimize the retrieval sequencing of drugs. For simplicity, the time taken for the robotic arm to load/unload a drug bin at the I/O point or storage location is ignored, as well as the time spent by the robotic arm moving between two I/O points. The notations used in this section are defined as follows.

Indices

| $k$ | The drug index in a prescription order $k \in \{1,2,\ldots,K\}$ |
| --- | --- |
| $p$ | The index of I/O points |
| $l_p$ | The last drug location index in the prior prescription order |
| $M_k$ | The location set index of bins of the drug $k$ |
| $\Phi$ | $\Phi = M_1 \cup M_2 \cup \cdots \cup M_K$ |
| $L$ | $L = \Phi \cup \{l_p\}$ |
| $i$ | The location index $i \in L$ |
| $j$ | The location index $j \in L$ |

Parameters

| $s_i$ | Remaining stock at the location $i$ |
| --- | --- |
| $X$ | Stochastic service time of pharmacists at the I/O point |
| $q_k$ | The dosage of the drug $k$ in the prescription order |
| $t_{i,j}$ | The travel time of the robotic arm on the route $O_p \to i \to j \to O_p$, which can be expressed by Eq.(1) |

Decision variables

| $x_{i,j}^p$ | $x_{i,j}^p = \begin{cases} 1, & \text{if the robotic arm travels along the route } O_p \to i \to j \to O_p \\ 0, & \text{otherwise} \end{cases}$ |
| --- | --- |
| $z_k^p$ | $z_k^p = \begin{cases} 1, & \text{if } l_p \in M_k \\ 0, & \text{if } l_p \notin M_k \end{cases}$ |

## 4.1. Retrieval sequencing model for layout A

To illustrate the operations of ADDSs with layout A, we consider a retrieval prescription order consisting of drugs $1,2,3,\ldots,K-1,K$. The robotic arm retrieves the drug bins with sequences $i_1, i_2, i_3, \ldots, i_{K-1}, i_K$, based on dual command operations, where $i_k$ denotes the location of drug $k$. As shown in Fig. 8, this is an example of picking a route under the scenario of successive order arrivals. There are two I/O points, i.e., $O_1$ and $O_2$. The picking route is $O_1 \to l_1 \to i_1 \to O_1$, $O_2 \to l_2 \to i_2 \to O_2$, $O_1 \to i_1 \to i_3 \to O_1$, $O_2 \to i_2 \to i_4 \to O_2$, L, $O_1 \to i_{K-3} \to i_{K-1} \to O_1$, $O_2 \to i_{K-2} \to i_K \to O_2$. $l_1$ and $l_2$ represent the last two drug locations from the prior prescription order. These two drugs need to be returned, and then the drugs from the new order are retrieved, forming a complete dual-command cycle.

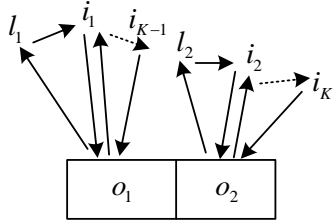

**Fig. 8**. Picking route by dual command operations for layout A.

To optimize the picking sequence of the prescription orders, a 0-1 integer programming model is established for ADDSs with layout A.

$$\min \sum_{p=1}^{2} \sum_{i=1}^{|L|} \sum_{\substack{j=1 \\ j \neq i}}^{|L|} E\left[\max\left(X, t_{i,j}\right)\right] x_{i,j}^{p} \tag{2}$$

subject to

$$q_k x_{i,j}^p \leq s_j, p \in \{1,2\}, i,j \in L, k \in \{1,2,\text{K},K\} \tag{3}$$

$$\sum_{j \in \Phi} \sum_{p \in \{1,2\}} x_{l_p,j}^p = 1 \tag{4}$$

$$\sum_{i \in L} \sum_{p \in \{1,2\}} x_{i,j}^p \leq 1, \forall j \in L \tag{5}$$

$$\sum_{j \in L} \sum_{p \in \{1,2\}} x_{i,j}^p \leq 1, \forall i \in L \tag{6}$$

$$\sum_{i \in L} \sum_{p \in \{1,2\}} x_{i,j}^p = \sum_{i \in L} \sum_{p \in \{1,2\}} x_{j,i}^p, \forall j \in L \tag{7}$$

$$\sum_{\substack{j \in M_k \\ j \neq i}} \sum_{i \in M_k} \sum_{p \in \{1,2\}} x_{i,j}^p = 0, \forall k \in \{1,2,\ldots,K\} \tag{8}$$

$$\sum_{\substack{j \in \Phi \\ j \notin M_k}} \sum_{i \in M_k} \sum_{p \in \{1,2\}} x_{i,j}^p = 1, \forall k \in \{1, 2, \ldots, K\} \tag{9}$$

$$\left| \sum_{\substack{j \in L \\ j \neq i}} \sum_{i \in L} x_{i,j}^1 - \sum_{\substack{j \in L \\ j \neq i}} \sum_{i \in L} x_{i,j}^2 \right| \leq 1 \tag{10}$$

$$\sum_{k=1}^{K} \sum_{p \in \{1,2\}} z_k^p \leq 2 \tag{11}$$

$$\sum_{\substack{j \in L \\ j \neq i}} \sum_{i \in L} \sum_{p \in \{1,2\}} x_{i,j}^p = K + 2 - \sum_{k=1}^{K} \sum_{p \in \{1,2\}} z_k^p \tag{12}$$

$$u_i - u_j + |L| x_{i,j}^p \leq |L| - 1, \forall 1 < i \neq j \leq |L|, p \in \{1, 2\} \tag{13}$$

$$x_{i,j}^p = 0, 1, \forall \, i, j \in L, i \neq j, p \in \{1, 2\} \tag{14}$$

$$z_k^p = 0, 1, \forall k \in \{1, 2, \ldots, K\}, p \in \{1, 2\} \tag{15}$$

The objective function (2) is to minimize the average picking time for the order picking in a human-machine cooperation environment, where

$$E\left[\max\left(X, t_{i,j}\right)\right] = t_{i,j} \int_0^{t_{i,j}} f(x) dx + \int_{t_{i,j}}^{+\infty} x f(x) dx. \tag{16}$$

Constraint (3) ensures that the remaining stock at location $j$ meets the dosage of the drug $k$. Eq.(4) indicates that the robotic arm returns the drug bin from the I/O point to the storage location $l_p$ and then retrieves the drug bin at location $j$. Constraints (5) and (6) limit that each drug location can be visited at most once. Eq.(7) indicates that the number of input arcs at each point is equal to the number of output arcs. Eq.(8) avoids the drug being repeatedly picked. Eq.(9) ensures that the robotic arm operates from one drug set to another drug set. The constraint (10) guarantees that the difference in the number of dual command operations initiated by the robotic arm from the two I/O points does not exceed 1, as the drugs are alternately picked from the two I/O points. Constraint (11) means that a new prescription order contains a maximum of two types of drugs that are the same as those in the prior prescription order. If the new prescription includes the last two drug requests in the prior prescription order, the pharmacist will sort it directly at the I/O point. Eq.(12) ensures that there are $K+2-\sum_{k=1}^{K}\sum_{p\in\{1,2\}}z_k^p$ dual command cycle operations. The sub-tour elimination constraints are described by Eq.(13), where $u_i$ are arbitrary real numbers (Miller et al., 1960).

### 4.2. Retrieval sequencing model for layout B

The operations of ADDSs with layout B are all dual command cycles. The cooperation between the robotic arm and the pharmacist is sequential. For the same retrieval sequence in Fig. 8, we show an example of the picking route under layout B. There is only one I/O point, i.e., $O$. The picking route is $O \to l_1 \to i_1 \to O$, $O \to i_1 \to i_2 \to O$, …, $O \to i_{K-1} \to i_K \to O$. $l_1$ represents the last drug location from the prior prescription order. This drug needs to be returned, and then the drugs from the new order are retrieved, forming a complete dual-command cycle.

To optimize the picking sequence of $K$ drugs, a 0-1 integer programming model is established for ADDSs with layout B.

$$\min \sum_{i=1}^{|L|} \sum_{\substack{j=1 \\ j \neq i}}^{|L|} \left[ x_{i,j}^1 t_{i,j} + x_{i,j}^1 E(X) \right] \quad (17)$$

subject to

$$q_k x_{i,j}^1 \leq s_j, \forall i,j \in L, k \in \{1,2,\ldots,K\} \quad (18)$$

$$\sum_{j \in \Phi} x_{l_1,j}^1 = 1, \quad (19)$$

$$\sum_{i \in L} x^1_{i,j} \leq 1, \forall j \in L \tag{20}$$

$$\sum_{j \in L} x^1_{i,j} \leq 1, \forall i \in L \tag{21}$$

$$\sum_{i \in L} x^1_{i,j} = \sum_{i \in L} x^1_{j,i}, \forall j \in L \tag{22}$$

$$\sum_{\substack{j \in M_k \\ j \neq i}} \sum_{i \in M_k} x^1_{i,j} = 0, \forall k \in \{1,2,\ldots,K\} \tag{23}$$

$$\sum_{j \notin M_k} \sum_{i \in M_k} x^1_{i,j} = 1, \forall k \in \{1,2,\ldots,K\} \tag{24}$$

$$\sum_{k=1}^{K} z^1_k \leq 1 \tag{25}$$

$$\sum_{\substack{j \in L \\ j \neq i}} \sum_{i \in L} x^1_{i,j} = K + 1 - \sum_{k=1}^{K} z^1_k \tag{26}$$

$$u_i - u_j + |L| x^1_{i,j} \leq |L| - 1, \forall 1 < i \neq j \leq |L| \tag{27}$$

$$x^1_{i,j} = 0,1, \forall\, i,j \in L, i \neq j \tag{28}$$

$$z^1_k = 0,1, \forall k \in \{1,2,\ldots,K\} \tag{29}$$

The objective function (17) aims to minimize the average picking time for order picking of ADDSs within a human-machine cooperation environment. The constraints (18)-(29) are nearly the same as constraints (3)-(9) and (11)-(15), and the explanation is omitted here.

The 0-1 integer programming models in Sections 4.1 and 4.2 can be solved by the Yalmip toolkit and Cplex solver.

## 5. Numerical Experiments

In this section, we evaluate the performance of ADDSs with layouts A and B by the average picking time of prescription orders. To demonstrate improvements in the picking efficiency of ADDSs, we compare the proposed method with other commonly used strategies that are described in detail in section 4.2.

We use the prescription order data set provided by Changi General Hospital in Singapore and randomly select 100 prescription orders. Each ADDS is equipped with a double-side storage rack and there are $M \times N$ ($M = 17$, $N = 17$) positions to locate bins on each side. For layout A, the I/O points are located at positions $(10, 9)$ and $(10, 10)$ on side 1. For layout B, position $(10, 9)$ on side 1 is an I/O point. The height of the bin is $d_m = 0.275m$, and the length is $d_n = 0.168m$. The robotic arm can move at the same speed $v$ in horizontal and vertical directions simultaneously, where $v = 0.1486ms^{-1}$.

**5.1. Performance comparison for layout A and layout B**

In this section, we examine the average picking time of 100 prescription orders in the human-machine cooperation environment, comparing layouts A and B. Let the mean $\mu$ be 5, 10, 15, 20, 25 and standard deviation $\sigma$ be 0, 5, 10, 15. Let $T_A$ and $T_B$ be the average picking time of prescription orders with layouts A and B respectively. We quantify the improvement in picking efficiency as $(1/T_A - 1/T_B)/(1/T_B)$. The experimental results are shown in Table 1.

Table 1

Average picking time for different parameters of 'human' under layout A and layout B.

| $(\mu, \sigma)$ | Layout A | Layout B | Improved |
|---|---|---|---|
| (5,0) | 88.27 | 124.82 | 41.40% |
| (5,5) | 91.34 | 124.82 | 36.65% |
| (5,10) | 100.93 | 124.82 | 23.67% |
| (5,15) | 113.02 | 124.82 | 10.43% |
| (10,0) | 94.98 | 162.22 | 70.78% |
| (10,5) | 101.31 | 162.22 | 60.12% |
| (10,10) | 113.74 | 162.22 | 42.62% |
| (10,15) | 127.45 | 162.22 | 27.28% |
| (15,0) | 116.67 | 199.62 | 71.10% |
| (15,5) | 122.19 | 199.62 | 63.37% |
| (15,10) | 133.39 | 199.62 | 49.65% |
| (15,15) | 146.40 | 199.62 | 36.34% |
| (20,0) | 148.72 | 237.02 | 59.38% |
| (20,5) | 151.41 | 237.02 | 56.54% |
| (20,10) | 159.12 | 237.02 | 48.96% |
| (20,15) | 170.08 | 237.02 | 39.36% |
| (25,0) | 184.13 | 274.42 | 49.04% |
| (25,5) | 186.26 | 274.42 | 47.34% |
| (25,10) | 189.43 | 274.42 | 44.87% |
| (25,15) | 197.78 | 274.42 | 38.75% |

Table 1 presents the results indicating that layout B yields unsatisfactory results. As expected, layout A demonstrates a decrease in the average picking time for prescription orders. Specifically, layout A shows an average picking time improvement of 10.43% to 71.10% compared to layout B. This improvement is primarily due to the synchronized service between the pharmacist and the robotic arm for the ADDS with layout A. When the robotic arm retrieves a new drug bin, the pharmacist sorts another type of drug at the I/O point. This synchronization significantly saves the average picking time for prescription orders in ADDSs with layout A.

In Table 1, the average picking time of layout A is related to $\mu$ and $\sigma$. The human-machine cooperation is a synchronous operation in layout A and the service time of the pharmacist influences the average picking time, which is determined by Eq.(16). With the increase of $\sigma$, it is more possible that the service time of pharmacists at the I/O point is longer than the robotic arm's retrieval time which causes the ADDS to wait for the pharmacists. While in layout B, the human-machine cooperation is a sequential operation and only the mean value of the service time for pharmacists influences the average picking time. The average picking time of layout B is fixed for specified $\mu$ and it doesn't depend on $\sigma$.

We illustrate the variation in picking efficiency improvement with $\mu$, where $\mu$ ranges from 1 to 50. Additionally, we set $\sigma$ as a constant, taking values of 0, 5, 10, and 15, respectively. The results are shown in Fig. 9.

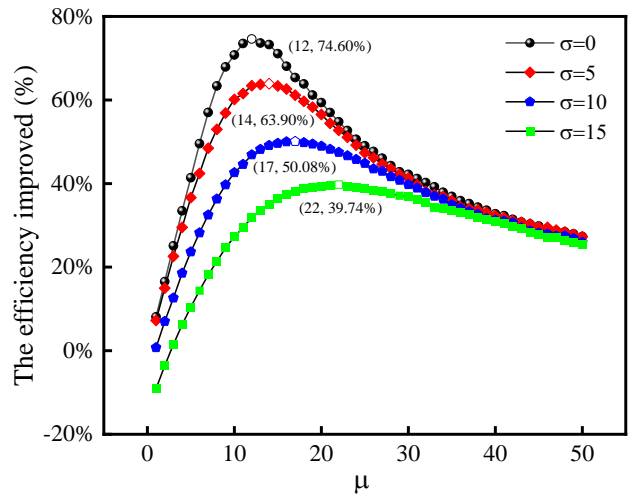

**Fig. 9**. The picking efficiency improved by layout A compared with layout B.

In Fig. 9, we compare the improved picking efficiency of layout A for various $\mu$ and $\sigma$. Overall, the picking efficiency is improved under most circumstances. This is mainly because, under the layout design of two I/O points, the operations of pharmacists and the robotic arm of ADDSs are parallel, involving parallel cooperation between the pharmacist and the robotic arm. In contrast, under the layout design of a single I/O point, the operations of the pharmacist and the robotic arm are sequential. The picking efficiency improvement is significantly higher in Layout A compared to Layout B. For a fixed $\sigma$, the improvement of picking efficiency in layout A compared to layout B exhibits a concave trend with respect to the $\mu$ value. Specifically, when

$\sigma$ is 0, 5, 10, and 15, the most significant efficiency improvements are observed at $\mu$ values of 12, 14, 17, and 22, respectively. The efficiency improvements are 74.60%, 63.9%, 50.08%, and 39.74% respectively.

**5.2. Retrieval sequencing strategies comparison**

In this section, we do a comparative analysis of the average prescription order picking time between the proposed optimal retrieval sequencing method and three common methods: dynamic programming, greedy, and random strategies. The details of the four picking strategies are described as follows.

**Dynamic programming.** The retrieval sequence follows the original drug sequence specified in the prescription order, and the locations of all bins for each drug are considered as a stage. At each stage, the storage locations of the drug serve as the states. The decision-making task at each stage involves selecting the appropriate drug bin to be picked. The dynamic programming approach aims to identify an optimal strategy that minimizes the overall picking time for a given prescription order, based on the sequence of stages (i.e., drugs).

**Greedy method**. The picking process follows the drug sequence specified in the prescription order. At each stage, the drug location closest to the current state is determined for the subsequent stage. However, the greedy method only focuses on optimizing the local picking time at each stage and does not consider the overall picking time for the prescription order.

**Random method.** The picking process follows the drug sequence specified in the prescription order. Following the determination of the drug location for the current stage, the drug location for the next stage is selected randomly. In contrast, the random method randomly selects drugs without considering any specific criteria or optimization objectives.

**Optimum Method.** The optimum retrieval sequencing method introduced in Section 4 not only optimizes the sequence of drugs (stages) but also takes into account the optimal selection of drug bin locations (states).

To compare the above strategies, we conducted numerical experiments in diverse human-machine cooperation environments. Specifically, we considered values of $\mu$ as 5, 10, and 25, and values of $\sigma$ as 5 and 10. The results of the four picking strategies for both layout A and layout B designs are depicted in Fig. 10.

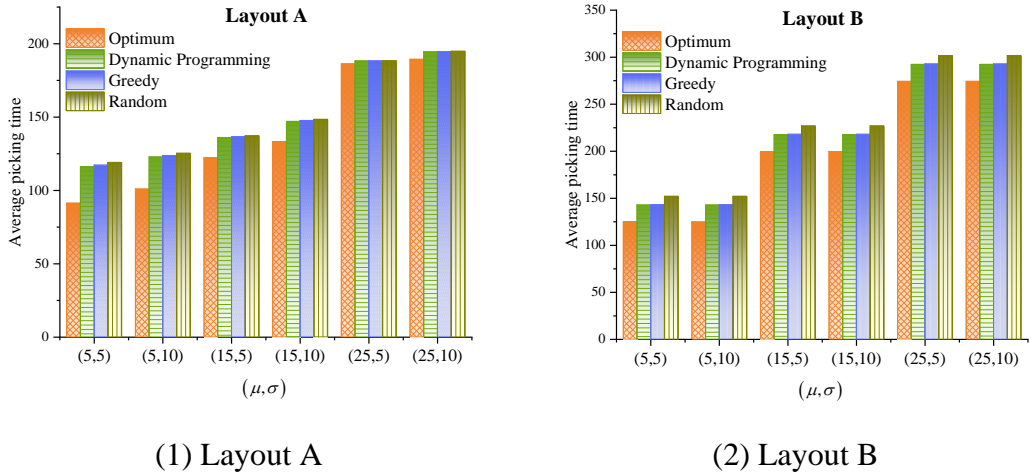

(1) Layout A          (2) Layout B

**Fig. 10.** The average picking time of the four picking strategies under different layout designs.

The observations in Fig. 10 can be summarized as follows.

**Observation 1:** Among the four strategies, the average picking time follows the ascending order of optimum < dynamic programming < greedy < random. The optimum method shows the lowest average picking time for all scenarios. This is because the optimum method takes into account the sequence of both drug types and drug bins. In contrast, the other three methods do not optimize the sequence of drug types, resulting in less efficient picking times. Notably, the average picking time for all four strategies in layout A is significantly lower compared to layout B, which is consistent with the findings in Section 5.1. This is mainly due to the cooperation time between the pharmacist and the robotic arm of the ADDS in Layout A.

**Observation 2:** For layout A, as $\mu$ and $\sigma$ increase, the difference between the average picking time under the optimum method and the other three strategies decreases. This can be primarily attributed to the fact that when both $\mu$ and $\sigma$ are large, there is a higher likelihood of the robotic arm having to wait for the pharmacist, leading to the 'human' becoming a bottleneck in the human-machine cooperation process. Similar conclusions can be drawn for layout B, but the magnitude of difference in the average picking time among the four strategies remains the same for the fixed $\mu$ and various $\sigma$. This is because layout B does not involve human-machine cooperation, and the human and machine operate sequentially. As a result, the four strategies in layout B are less influenced by the 'human'. Additionally, the average picking time in layout B is unaffected by the standard deviation $\sigma$.

## 6. Conclusion

In this paper, we analyze the performance of automated drug dispensing systems (ADDSs) in a human-machine cooperation environment under different layout designs. The retrieval sequencing models for two layout designs are established to optimize the retrieval sequence of drugs for successive prescription order arrivals. Numerical experiments are conducted, and the results indicate that the performance of the ADDS with two I/O points (layout A) is superior to the layout with one I/O point (layout B). Additionally, for different system layouts, we find that the variation in the stochastic sorting time of 'human' has an impact on system efficiency improvement. Under different means and standard deviations, the improvement in system efficiency follows a concave function. This finding provides valuable insights for pharmacy technicians. We compare the proposed strategies with dynamic programming, greedy, and random strategies, and the results demonstrate the advantage of our approach in improving the average picking time of the system.

  Note that this paper focuses on optimizing the retrieval sequence of drugs in prescription orders, following the common practice of first-come, first-served strategy in pharmacies. In future research, batch service rules, priority service rules, and other service rules can be analyzed for the ADDS. Furthermore, while this study considers the 'human' sorting time to follow a normal distribution, it would be interesting to explore the impact of other distributions on the average picking time of the system. It is important to acknowledge that our optimization efforts in layout design for ADDSs have primarily focused on system configurations with one or two I/O points. However, it is crucial to recognize that modelling for layouts with multiple I/O points introduces increased complexity. This complexity poses an important challenge. In order to fully understand the impact of different I/O point quantities on system performance and efficiency, future studies should delve into the intricate modelling aspects of ADDS layouts with multiple I/O points.

**Disclosure statement**

No potential conflict of interest was reported by the author(s).

**Data availability statement**

The data that support the findings of this study are available from the corresponding author upon reasonable request.

30**Reference**

Ardjmand, E., Shakeri, H., Singh, M.& Sanei Bajgiran, O. 2018. Minimizing order picking makespan with multiple pickers in a wave picking warehouse. Int. J. Prod. Econ., 206, 169-183. https://doi.org/10.1016/j.ijpe.2018.10.001.

Azadeh, K., De Koster, R.& Roy, D. 2019. Robotized and automated warehouse systems: Review and recent developments. Transp. Sci., 53(4), 917-945. https://doi.org/10.1287/trsc.2018.0873.

Azadeh, K., Roy, D., de Koster, R.& Khalilabadi, S. M. G. 2023. Zoning strategies for human–robot collaborative picking. Decis. Sci. https://doi.org/10.1111/deci.12620.

Bortolini, M., Galizia, F. G., Gamberi, M.& Gualano, F. 2020. Integration of single and dual command operations in non-traditional warehouse design. Int. J. Adv. Manuf. Tech., 111(9-10), 2461-2473. https://doi.org/10.1007/s00170-020-06235-4.

Boysen, N., Fedtke, S.& Weidinger, F. 2018. Optimizing automated sorting in warehouses: The minimum order spread sequencing problem. Eur. J. Oper. Res., 270(1), 386-400. https://doi.org/10.1016/j.ejor.2018.03.026.

Boysen, N.& Stephan, K. 2016. A survey on single crane scheduling in automated storage/retrieval systems. Eur. J. Oper. Res., 254(3), 691-704. https://doi.org/10.1016/j.ejor.2016.04.008.

Chen, T.-L., Cheng, C.-Y., Chen, Y.-Y.& Chan, L.-K. 2015. An efficient hybrid algorithm for integrated order batching, sequencing and routing problem. Int. J. Prod. Econ., 159, 158-167. https://doi.org/10.1016/j.ijpe.2014.09.029.

Chen, W., De Koster, R.& Gong, Y. 2023. Warehouses without aisles: Layout design of a multi-deep rack climbing robotic system. Transportation Research Part E: Logistics and Transportation Review, 179. https://doi.org/10.1016/j.tre.2023.103281.

da Costa Barros, Í. R.& Nascimento, T. P. 2021. Robotic mobile fulfillment systems: A survey on recent developments and research opportunities. Rob. Auton. Syst., 137. https://doi.org/10.1016/j.robot.2021.103729.

De Koster, R., Le-Duc, T.& Roodbergen, K. J. 2007. Design and control of warehouse order picking: A literature review. Eur. J. Oper. Res., 182(2), 481-501. https://doi.org/10.1016/j.ejor.2006.07.009.

De Lombaert, T., Braekers, K., De Koster, R.& Ramaekers, K. 2022. In pursuit of humanised order picking planning: Methodological review, literature classification and input from practice. Int. J. Prod. Res., 61(10), 3300-3330. https://doi.org/10.1080/00207543.2022.2079437.

Gu, J., Goetschalckx, M.& McGinnis, L. F. 2010. Research on warehouse design and performance evaluation: A comprehensive review. Eur. J. Oper. Res., 203(3), 539-549. https://doi.org/10.1016/j.ejor.2009.07.031.

Khader, N., Lashier, A.& Yoon, S. W. 2016. Pharmacy robotic dispensing and planogram analysis using association rule mining with prescription data. Expert Syst. Appl., 57(Sep.), 296-310. https://doi.org/10.1016/j.eswa.2016.02.045.

Li, D., Ruan, X.& Yue, Q. 2022. Optimization study of three-stage assembly flowshop problem in pharmacy automation dispensing systems. Comput. Oper. Res., 144(Aug.), 105810. https://doi.org/10.1016/j.cor.2022.105810.

Lin, C.-Y.& Hsieh, P.-J. 2017. Development of an automatic dispensing system for traditional chinese herbs. J. Healthcare Eng., 2017, 12. https://doi.org/10.1155/2017/9013508.

Löffler, M., Boysen, N.& Schneider, M. 2023. Human-robot cooperation: Coordinating autonomous mobile robots and human order pickers. Transp. Sci. https://doi.org/10.1287/trsc.2023.1207.

Mahajan, S., Rao, B. V.& Peters, B. A. 1998. A retrieval sequencing heuristic for miniload end-of-aisle automated storage/retrieval systems. Int. J. Prod. Res., 36(6), 1715-1731. https://doi.org/10.1080/002075498193246.

Miller, C. E., Tucker, A. W.& Zemlin, R. A. 1960. Integer programming formulation of traveling salesman problems. J. ACM, 7(4), 326–329. https://doi.org/10.1145/321043.321046.

Pasparakis, A., De Vries, J.& De Koster, R. 2023. Assessing the impact of human–robot collaborative order picking systems on warehouse workers. Int. J. Prod. Res., 61(22), 7776-7790. https://doi.org/10.1080/00207543.2023.2183343.

Pohl, L. M., Meller, R. D.& Gue, K. R. 2009. An analysis of dual-command operations in common warehouse designs. Transp. Res. E, 45(3), 367-379. https://doi.org/10.1016/j.tre.2008.09.010.

Randhawa, S. U., Mcdowell, E. D.& Wang, W.-T. 1991. Evalution of scheduling rules for single-and dual-dock automated storage/retrieval system. Comput. Ind. Eng., 20(4), 401-410. https://doi.org/10.1016/0360-8352(91)90012-U.

Randhawa, S. U.& Shroff, R. 1995. Simulation-based design evalution of unit load automated storage/retrieval systems. Comput. Ind. Eng., 28(1), 71-79. https://doi.org/10.1016/0360-8352(94)00027-k.

Roodbergen, K. J.& Vis, I. F. A. 2009. A survey of literature on automated storage and retrieval systems. Eur. J. Oper. Res., 194(2), 343-362. https://doi.org/10.1016/j.ejor.2008.01.038.

Salah, B., Janeh, O., Noche, B., Bruckmann, T.& Darmoul, S. 2017. Design and simulation based validation of the control architecture of a stacker crane based on an innovative wire-driven robot. Rob. Comput. Integr. Manuf., 44, 117-128. https://doi.org/10.1016/j.rcim.2016.08.010.

van Gils, T., Ramaekers, K., Caris, A.& de Koster, R. B. M. 2018. Designing efficient order picking systems by combining planning problems: State-of-the-art classification and review. Eur. J. Oper. Res., 267(1), 1-15. https://doi.org/10.1016/j.ejor.2017.09.002.

Wang, Z., Sheu, J. B., Teo, C. P.& Xue, G. 2022. Robot scheduling for mobile‐rack warehouses: Human–robot coordinated order picking systems. Prod. Oper. Manag., 31(1), 98-116. https://doi.org/10.1111/poms.13406.

Wauters, T., Villa, F., Christiaens, J., Alvarez-Valdes, R.& Vanden Berghe, G. 2016. A decomposition approach to dual shuttle automated storage and retrieval systems. Comput. Ind. Eng., 101, 325-337. https://doi.org/10.1016/j.cie.2016.09.013.

Yang, J., de Koster, R. B. M., Guo, X.& Yu, Y. 2023. Scheduling shuttles in deep-lane shuttle-based storage systems. Eur. J. Oper. Res., 308(2), 696-708. https://doi.org/10.1016/j.ejor.2022.11.037.

Yang, X., Hua, G., Hu, L., Cheng, T. C. E.& Huang, A. 2021. Joint optimization of order sequencing and rack scheduling in the robotic mobile fulfilment system. Comput. Oper. Res., 135. https://doi.org/10.1016/j.cor.2021.105467.

Yu, Y.& De Koster, R. B. M. 2012. Sequencing heuristics for storing and retrieving unit loads in 3d compact automated warehousing systems. IIE Trans., 44(2), 69-87. https://doi.org/10.1080/0740817x.2011.575441.

Yuan, M., Zhao, N., Wu, K.& Chen, Z. 2023. The storage location assignment problem of automated drug dispensing machines. Comput. Ind. Eng., 184, Article 109578. https://doi.org/10.1016/j.cie.2023.109578.

Zhang, J., Zhang, N., Tian, L., Zhou, Z.& Wang, P. 2023. Robots' picking efficiency and pickers' energy expenditure: The item storage assignment policy in robotic mobile fulfillment system. Comput. Ind. Eng., 176, Article 108918. https://doi.org/10.1016/j.cie.2022.108918.

Zhen, L.& Li, H. 2022. A literature review of smart warehouse operations management. FEM, 9(1), 31-55. https://doi.org/10.1007/s42524-021-0178-9.